\documentstyle[12pt,psfig]{article}
\begin{document}
\pagestyle{plain}
\setcounter{page}{1}
\baselineskip16pt

\begin{titlepage}

\begin{flushright}
PUPT-1708\\
hep-th/9707167
\end{flushright}
\vspace{20 mm}

\begin{center}
{\huge Strings and D-Branes at High Temperature}

\vspace{5mm}

\end{center}

\vspace{10 mm}

\begin{center}
{\large Sangmin Lee and L\'arus Thorlacius }

\vspace{3mm}

Joseph Henry Laboratories\\
Princeton University\\
Princeton, New Jersey 08544

\end{center}

\vspace{2cm}

\begin{center}
{\large Abstract}
\end{center}

The thermodynamics of a gas of strings and D-branes near the Hagedorn 
transition is described by a coupled set of Boltzmann equations for 
weakly interacting open and closed long strings. The resulting 
distributions are dominated by the open string sector, indicating 
that D-branes grow to fill space at high temperature.

\vspace{2cm}
\end{titlepage}
\newpage
\renewcommand{\baselinestretch}{1.1}  


\newcommand{\grad}{\nabla}
\newcommand{\tr}{\mathop{\rm tr}}
\newcommand{\half}{{1\over 2}}

\newcommand{\dint}[2]{\int_{#1}^{#2}}
\newcommand{\D}{\displaystyle}
\newcommand{\PDT}[1]{\frac{\partial #1}{\partial t}}
\newcommand{\PD}{\partial}
\newcommand{\newcaption}[1]{\centerline{\parbox{6in}{\caption{#1}}}}

\section{Introduction}

\noindent
Strings exhibit interesting behavior at high 
temperature.  A gas of closed fundamental 
strings approaching the Hagedorn temperature $T_H$ enters 
a long string phase, where any added energy goes into forming 
string rather than raising the temperature beyond $T_H$.   
In the long string phase the average number of long strings of 
a given length $\ell$ is proportional to $1/\ell$ and 
the average total number of long strings is $\log(E)$, 
where $E$ is the total energy in the microcanonical 
ensemble.  These results are readily found by solving 
a Boltzmann equation for long closed 
strings \cite{salska,lowtho}.  They can also be obtained 
from the exponentially rising density of states for
free strings, provided one is careful to properly define the 
microcanonical ensemble \cite{salska,bravaf,djnt}.
At very high energy density the description 
in terms of weakly coupled strings in flat spacetime breaks 
down and the physics of the system is unknown.  

Strings are not the only extended objects in string theory 
and it is natural to ask how the above physical picture is 
changed when the system also contains p-branes.  
We will address this question in the context of Type~II 
strings and D-branes, by extending the Boltzmann equation 
approach of ref.~\cite{lowtho} to include the auxiliary open 
string sector that describes the D-brane dynamics.  We find 
that the long closed string phase is suppressed in the 
presence of a dilute gas of D-branes (and anti-branes) and
the system is instead dominated by long open strings as the 
Hagedorn temperature is approached.  

The interactions of fundamental strings always include 
gravity and the thermodynamic limit, where one considers a fixed
energy density as the volume becomes arbitrarily large, is not
well defined for gravitational systems due to the Jeans instability.
For a fixed value of the string coupling $g$ we can only 
consider a thermal ensemble on finite length scales,
\begin{equation}
R^2 < {1\over g^2 \rho},
\label{jeans}
\end{equation}
where $\rho$ is the energy density.  Alternatively, the 
limit of large volume can be taken for fixed $\rho$ only 
if $g$ is simultaneously scaled to zero fast enough.

This has important 
consequences if we wish to include D-branes 
in the system.  A Dirichlet p-brane carries a conserved 
charge that couples to a Ramond-Ramond, rank p+1, 
antisymmetric gauge potential.  A finite volume target 
space must carry zero net charge, and as the D-branes 
are the only objects in the theory that carry the R-R 
charge, we must have an equal number of branes
and anti-branes in our system.\footnote{Some of the 
p-branes may be in bound states with (p+2)-branes or 
(p+4)-branes but the net p-brane charge
still has to be zero for all p.}
The annihilation of 
branes and anti-branes will deplete their numbers in a true
equilibrium configuration unless the energy density
is sufficiently high, $\rho \sim 1/g$, to have brane 
anti-brane pairs forming at a rapid rate.   At such high 
energy densities, however, the dynamics of 
D-branes are unknown and our approximations break 
down.  

We will instead consider a dilute collection of 
branes and anti-branes interacting with a gas of strings.
This is not an equilibrium configuration but if the 
average time between brane anti-brane encounters
is long compared to the time it takes the string gas to
reach equilibrium, then the configurations at intermediate
time scales will be well approximated by solutions to our
transport equations.
Note that we cannot take the D-branes to be too dilute.  
At any fixed value of the string coupling, the 
Jeans instability limits the size of the entire system and 
we want the average separation between branes to
be small on that scale.  As we shall see below, for weakly 
coupled strings there is a range of D-brane number densities 
where both of these requirements are satisfied.

\section{Boltzmann equation for closed strings}

We use Boltzmann transport equations to obtain the 
distribution of string lengths, both for the closed strings
in the bulk of spacetime and the open string sector 
associated to the D-branes.  Boltzmann equations for strings 
have previously been studied in a discrete model by Salomonson 
and Skagerstam \cite{salska}, and in a continuum approach by
Lowe and Thorlacius \cite{lowtho} which can be easily 
adapted to account for the two coupled string sectors.

To review, let us first consider the simpler system of only 
closed strings and no D-branes.  The Boltzmann equation
for weakly coupled closed strings is \cite{lowtho},
\begin{eqnarray}
{{\partial f(\ell)} \over {\partial t}}
&=& {\kappa\over V} \Bigl\{ -{1\over 2}\ell^2 f(\ell)
+ {1\over 2}\int_0^{\ell} d \ell'~
\ell' (\ell-\ell') f(\ell') f(\ell-\ell') 
- \ell f(\ell) \int_0^{\infty} d \ell' \ell' f(\ell') 
\nonumber\\
&&\qquad +
\int_0^{\infty} d \ell' (\ell+\ell') f(\ell+\ell') \Bigr\},
\label{loope}
\end{eqnarray}
where $f(\ell)$ is the average number of strings of length $\ell$,
$\kappa$ is a positive constant proportional to $g^2$, and
$V$ is the volume of the system.\footnote{We have
set $\alpha'=1$ for convenience.}  The terms on the right hand
side all involve the three closed string interaction shown in 
Figure~1a.  Higher order interactions of four or more closed 
strings may be ignored at weak string coupling.

The first term corresponds to a loop of length $\ell$ self-intersecting
and splitting into two strings of length $\ell'$ and $\ell-\ell'$, while 
the second term describes the reverse process where two smaller
strings join to form a string of length $\ell$.  The third term is due
to strings of length $\ell$ and $\ell'$ joining to form a single string
of length $\ell+\ell'$, while the fourth term represents a string of 
length $\ell+\ell'$ splitting into strings of length $\ell$ and $\ell'$.

In each of the four terms, two string segments meet and exchange
ends at the intersection point.  The interaction rate is proportional
to the coefficient $\kappa$ which involves
a sum over relative momentum and orientation of the segments.  In
the long string limit this sum does not depend on string length nor 
on whether the two segments belong to separate strings or the same
one, and thus the coefficient is the same for all four terms.  The 
overall factor of $1/V$ reflects the fact that the string segments have 
to be at the same place in the embedding space in order for an 
interaction to take place.

\begin{figure}
\psfig{file=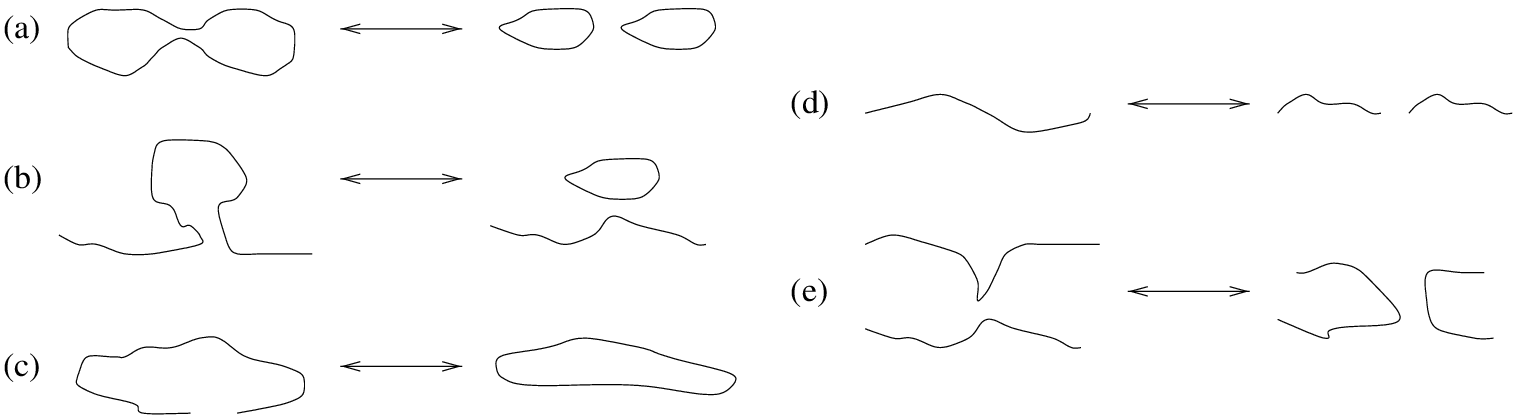}
\newcaption{Leading order string interactions. \label{hotfig}}
\end{figure}

At equilibrium we are interested in a static solution of the Boltzmann
equation, where ${\partial  \over \partial t}f(\ell)=0$, and the terms
on the right hand side of (\ref{loope}) cancel.  The equilibrium
solution is easily obtained \cite{lowtho},
\begin{equation}
f(\ell) = {1\over \ell} e^{-\ell/ L} \>,
\label{eqsol}
\end{equation}
where $L$ is the average total length of string in the ensemble.

The energy of a long string is to a good approximation proportional
to its length, $\varepsilon = \sigma\ell$, where $\sigma$ is the string
tension ($\sigma$ is $O(1)$ in our units).  
The distribution $f(\ell)$ is 
related to the single-string density of states $\omega(\varepsilon)$
as follows:
\begin{equation}
f(\ell) \,d\ell=  \omega(\varepsilon) e^{-\beta \varepsilon}
\, d\varepsilon.
\end{equation}
Inserting the equilibrium solution (\ref{eqsol}) we read off
\begin{equation}
\omega(\varepsilon) = { {e^{\beta_H \varepsilon}} \over \varepsilon},
\label{bdos}
\end{equation}
where 
\begin{equation}
\beta_H = \beta - {1 \over  L} \,,
\label{cetemp}
\end{equation}
is identified as the inverse Hagedorn temperature.  
The corresponding multi-string density of states is 
\begin{equation}
\Omega(E)=\exp (\beta_H E), 
\end{equation}
and the single-string distribution
function $d(\varepsilon,E)$, which gives the average number of
strings carrying energy between $\varepsilon$ and 
$\varepsilon+d\varepsilon$ in a system of total energy $E$, is
given by
\begin{equation}
d(\varepsilon, E)  \approx {{\omega(\varepsilon) \Omega(E-\varepsilon)}
\over {\Omega(E) }} \approx {1\over \varepsilon}.
\label{dftn}
\end{equation}
This result is valid for $c < \varepsilon < E-c$ where $c$ is some
constant independent of $E$ \cite{djnt}.

The average total number of long strings is then $\log E$.  At an 
energy density of order one in string units the average total length
of string in the gas is $L\sim V$ so the typical long string has
length
\begin{equation}
\bar\ell \sim {V\over \log V} \,,
\label{clength}
\end{equation}
which is large compared to the linear size of the system, 
$L_0 = V^{1/d}$.  

\section{Boltzmann equations for strings and D-branes}

We now include a dilute gas of D-branes in the system.
We consider parallel p-branes and anti-p-branes in 
a target space 
${\cal M}={\bf R}\otimes 
{\cal K}_\parallel \otimes {\cal K}_\perp$,
where ${\cal K}_\parallel$ and ${\cal K}_\perp$ have dimension 
$d_\parallel = p$ and $d_\perp = 9-p$ respectively.
For simplicity, we will assume $0\leq p \leq 6$.
The branes are wrapped around ${\cal K}_\parallel$, which
we will take to be compact with volume of order one in string units, 
but are free to move around in the transverse space ${\cal K}_\perp$.  
The transverse volume $V_\perp$ is macroscopic but bounded by the 
Jeans volume,
\begin{equation}
V_\perp < \left( {1\over g^2\rho}\right)^{d_\perp /2} \,.
\label{vperp}
\end{equation}
Let $n$ be the number density of p-branes and anti-branes in
the transverse space,
\begin{equation}
n = {N_p + \bar N_p \over V_\perp } = {2 N_p \over V_\perp} \,.
\label{ndens}
\end{equation}
The typical separation between branes is 
$l_p \sim n^{-1/d_\perp}$.  We must take this distance to
be large in string units for otherwise the energy density 
due to D-branes would no longer be small compared to $1/g$.

The string gas now contains both closed and open strings.
Let $f(\ell)$ denote the average number of closed strings 
of length $\ell$ in the system and $p(\ell)$ the average 
number of open strings of length $\ell$.  
There are two coupled Boltzmann equations which express the rate of 
change of $f(\ell)$ and $p(\ell)$ with time.  For weakly interacting 
strings the terms in the Boltzmann equations correspond to the
various string interactions shown in Fig.~\ref{hotfig}.
Let us consider the $f(\ell)$ equation first,
\begin{eqnarray}
 \D\PDT{f(\ell)} &=& 
{\kappa\over V} \Bigl\{ 
- \half \ell^2f(\ell)
+\half \dint{0}{\ell} d\ell'\ell'f(\ell')(\ell-\ell')f(\ell-\ell') 
- \ell f(\ell) \dint{0}{\infty} d\ell'\ell'f(\ell')  
\cr  & & \quad
+ \dint{0}{\infty} d\ell'(\ell+\ell')f(\ell+\ell')   \Bigr\} 
+ {b\over 2nV} p(\ell) - an \ell f(\ell)
\cr & &
+ \D{\kappa\over V} \Bigl\{ 
\dint{\ell}{\infty} d\ell'(\ell'-\ell)p(\ell') 
- \ell f(\ell) \dint{0}{\infty} d\ell'\ell'p(\ell') \Bigr\} .
\label{dfdt}
\end{eqnarray}
The first four terms on the right hand side correspond to the closed string 
interaction in Figure~1a and are the same as in the
Boltzmann equation of the previous section.  
The remaining terms correspond to closed strings interacting with 
open strings attached to D-branes, as indicated in Figure~1b and 1c.  
A splitting interaction where a closed string is converted into 
an open string, or an open string splits into two open strings, 
comes with a factor $a$, which is proportional to $g$ and includes 
a sum over the momentum and relative orientation of the string segments
involved.  The reverse joining process is characterized by another
coefficient $b$, which is also proportional to $g$.

The equation for $p(\ell)$ receives contributions from the open string 
interactions in Figure~1d and 1e, and also from the mixed interactions in
Figure~1b and 1c, 
\begin{eqnarray}
 \D\PDT{p(\ell)} &=& 
{b\over 2nV} \dint{0}{\ell} d\ell'p(\ell')p(\ell-\ell') 
- an \ell p(\ell) 
+ 2an\dint{\ell}{\infty}d\ell'p(\ell')
- {b\over nV}p(\ell)\dint{0}{\infty} d\ell'p(\ell') 
 \cr & &
+ \D{\kappa\over V}\Bigl\{
\D\dint{0}{\ell}d\ell_1 p(\ell_1) \dint{\ell-\ell_1}{\ell}d\ell_2 
(\ell_1+\ell_2-\ell) p(\ell_2)
+2\dint{0}{\ell}d\ell_1\ell_1 p(\ell_1)\dint{\ell}{\infty}d\ell_2p(\ell_2)
\cr & & \quad
+\D\ell\dint{\ell}{\infty}d\ell_1 p(\ell_1)\dint{\ell}{\infty}d\ell_2p(\ell_2) 
-\ell p(\ell)\dint{0}{\infty} d\ell'\ell'p(\ell') \Bigr\}
\cr & &
+ \D{\kappa\over V}\Bigl\{
\dint{0}{\ell} d\ell'\ell'f(\ell')(\ell-\ell')p(\ell-\ell') 
- \half \ell^2p(\ell)
+ \ell\dint{\ell}{\infty}d\ell'p(\ell')
\cr & & \quad
- \ell p(\ell)\dint{0}{\infty}d\ell'\ell'f(\ell') \Bigr\}
+ an \ell f(\ell) - \D{b\over 2nV} p(\ell). 
\label{dpdt}\end{eqnarray}
At equilibrium, the distribution of strings does not change with time,
\begin{equation}
\D\PDT{f} = \PDT{p} = 0.
\label{equil}\end{equation}

The coupled equations (\ref{dfdt}) and (\ref{dpdt}) appear considerably 
more complicated than the Boltzmann equation (\ref{loope}) for closed strings.
One nevertheless finds a simple equilibrium solution,
\begin{equation}
f(\ell)=\D{1\over \ell}\exp[-\ell/L_c],
\qquad   p(\ell)=\D{N_o\over L_c} \exp[-\ell/L_c].
\label{sol2}\end{equation}
The distributions are characterized by the average total length of closed and 
open string and the average total number of open strings,
\begin{equation}
L_c\equiv\dint{0}{\infty}d\ell\, \ell f(\ell),\qquad 
L_o\equiv \dint{0}{\infty}d\ell\, \ell p(\ell),\qquad
N_o\equiv \dint{0}{\infty}d\ell\, p(\ell). 
\label{parameters}\end{equation}
These parameters satisfy the relations
\begin{equation}
{N_o\over L_c} = \D{2a n^2 V\over b},\qquad N_o L_c = L_o . 
\label{cond}\end{equation}
It immediately follows that the open strings tend to dominate
the ensemble.  Whenever there is a macroscopic number of open 
strings present, $N_o >> 1$, the total length
of closed string will be vanishing compared to the 
total length of open string, $L_c << L_o$.
In other words, the D-branes efficiently chop up the strings so 
that the closed string gas is replaced by a web of open strings 
attached to the D-branes.  

The typical length of a long open string can be obtained from
the single string distribution function for open strings
$d_o(\varepsilon,E)$ in the microcanonical ensemble.
The first step is to read off the single-string density of 
states for open strings from the equilibrium solution
(\ref{sol2}),
\begin{equation}
\omega_o(\varepsilon)  
= {2an^2V\over b\sigma} \exp{(\beta_H\varepsilon)}.
\end{equation}
The corresponding multi-string density of states can be 
obtained in a saddle point approximation, which is valid when
the total number of D-branes in the gas is large,
\begin{equation}
\Omega_o(E) \approx \exp \Bigl\{
\sqrt{8an^2VE\over b\sigma} + \beta_H E \Bigr\}.
\end{equation}
The open string distribution function is then
\begin{eqnarray}
d_o(\varepsilon,E) 
&\approx & 
{\omega_o(\varepsilon)\Omega_o(E-\varepsilon)\over \Omega_o(E)}
\cr 
&\sim &
\exp \Bigl\{-\sqrt{8an^2VE\over b\sigma}
\bigl(1-\sqrt{1-{\varepsilon\over E}}\bigr)\Bigr\}.
\end{eqnarray}
The distribution function has a characteristic energy scale,
\begin{equation}
\varepsilon_0 = {1\over n}\sqrt{b\rho\over a\sigma},
\end{equation}
below which the distribution is more or less flat and above which it is
strongly suppressed.  The energy of a long string is proportional to 
its length so there is a corresponding characteristic length scale for
the open strings in the gas.  Since the rate coefficients $a$ and $b$
are both proportional to $g$ the characteristic length is given in 
string units by
\begin{equation}
l_0 \sim {\rho^{1/2}\over n} ,
\label{openlength}
\end{equation}
up to factors of order one.  At the Hagedorn energy density, $\rho\sim 1$,
the characteristic open string length is long compared to the typical D-brane
separation $l_p\sim n^{-1/d_\perp}$, so the D-branes are effectively 
overlapping.

Our calculations have involved a number of assumptions and we close 
this section with a discussion of their range of validity.  
The Boltzmann equations
only have the relatively simple form of (\ref{dfdt}) and (\ref{dpdt}) if 
we can assume that points on a string that are separated by a finite 
parameter distance are at uncorrelated positions in the embedding
space.  This criterion is satisfied if the volume $V_{s}$ occupied by
a long string when placed in an infinite space is large compared to the
system volume $V$.  The embedding of a long string of length $\ell$ 
is a random walk which occupies a volume 
$V_s \sim \ell^{d_\perp/2}$ in $d_\perp$ spatial dimensions.
Let us assume that the energy density in the system is order one in string
units and that the spatial volume is close to the maximum value allowed
before the Jeans instability leads to gravitational collapse,
$V\sim g^{-d_\perp}$.  The characteristic open string length
(\ref{openlength}) involves the D-brane number density and we find
that $V_s>V$ is satisfied only if 
\begin{equation}
n<g^2.
\label{ncond}
\end{equation}
We also want to impose the condition that the typical D-brane separation,
$l_p \sim n^{-1/d_\perp}$, is small compared to the linear size of the
system, which requires $n>g^{d_\perp}$.  Since $d_\perp\geq 3$ for
p-branes with $p\leq 6$ there is always an allowed range for the number
density.  

If $V_s$ for typical strings is smaller than the system volume the terms 
in the Boltzmann equations that correspond to string self-interactions 
have to be modified by replacing the factor of $V$ in the denominator
by $V_s$.  We have not found explicit solutions of the resulting 
equations but we expect such a modification to increase the weight of 
shorter strings in the distribution when the D-brane number density is
higher than in (\ref{ncond}).

Our results also rest on the assumption that the strings take less time to
reach an equilibrium configuration than the timescale on which the 
D-branes and anti-branes meet and annihilate.  It turns out that this is
a weaker condition than (\ref{ncond}).  The D-brane lifetime can be 
estimated as
\begin{equation}
\tau \sim {1\over n\sigma v},
\end{equation}
where $\sigma$ is the brane anti-brane annihilation cross section and
$v$ is the average D-brane speed.  The D-branes are immersed in a
string gas with an energy density of order one in string units.  Their 
kinetic energy will therefore be of order one which means that their 
motion is non-relativistic with $v\sim g^{1/2}$.
The eikonal approximation to the low velocity scattering amplitude of
a D-brane and anti-brane pair develops a tachyonic instability if the
impact parameter is less than a certain value of order one in 
string units \cite{bansus}.  We use this as an estimate for the 
annihilation cross section and take $\sigma\sim O(1)$.  The D-brane
lifetime is then found to be $\tau\sim g^{-1/2}n^{-1}$.

This is to be compared to the time it takes the string distribution to 
find equilibrium.  For an ensemble dominated by open strings the 
relevant processes are the open string interactions in Figure 1d and 1e
and we can ignore the closed strings.  In the long string limit the
probability for a pair of string segments, that happen to come across
each other, to interact is independent of the overall string lengths.
Furthermore, if a long string undergoes an interaction its new length
will be largely uncorrelated to the old one.  It follows that a gas 
of long strings will find equilibrium on a time scale where the strings
have interacted a few times each.  It turns out that the exchange 
interaction Figure 1e, which occurs in the bulk of spacetime, is more
efficient at redistributing length among the open strings than the 
endpoint interaction in Figure 1d, which can only take place inside
a D-brane worldvolume.  The probability for a particular string to 
participate in such an exchange interaction in a unit of time goes 
as $g^2 \ell$, where $\ell$ is the length of the string.  Using the
characteristic length (\ref{openlength}), and assuming $\rho\sim 1$,
the time to reach equilibrium is
\begin{equation}
\tau_e \sim {n\over g^2}.
\label{eqtime}
\end{equation}
It is short compared to the D-brane lifetime if the D-brane number
density satisfies $n<g^{3/4}$, which is a weaker condition than
(\ref{ncond}).  The equilibrium time scale $\tau_e$ can also be
estimated by linearizing the long string Boltzmann equation
(\ref{dpdt}) around the equilibrium solution (\ref{sol2}) and
find the rate at which the perturbation decays.  This second
method gives an answer that agrees with (\ref{eqtime}).

\section{Discussion}

D-branes strongly affect the long string phase associated with the Hagedorn
transition.  In the absence of D-branes it consists of long closed strings
that traverse the entire system many times over, but a dilute collection of
D-branes will suppress the long closed strings and replace them with long
open strings that are attached to the D-branes.  The typical open string is 
long compared to the D-brane separation and is therefore most likely to have
its ends on different D-branes.  Since open strings are associated to the 
dynamics of the D-branes themselves these results suggest a physical
picture where D-branes grow with temperature as the Hagedorn transition
is approached and eventually overlap to fill space.

The system enters the long string phase when the energy density becomes
of order one in string units.  For a D-brane number density that
satisfies the condition (\ref{ncond}) the `bare' energy associated 
with the D-branes is then small compared to the energy carried in the 
open strings attached to them.  If the energy density of the system is
increased further the characteristic open string length increases and 
the D-branes become more interconnected.  If the energy density becomes
of order $1/g$ non-perturbative effects, such as pair creation of 
branes and anti-branes, become important and our description in terms
of weakly interacting strings breaks down.

In this paper we took the string coupling to be weak and limited the
system size in order to avoid the Jeans instability but it would be 
interesting to incorporate long range gravitational effects into the
description of the string distributions.  Some related issues are
addressed in recent work of Horowitz and Polchinski \cite{horpol}.

\section*{Acknowledgements}

We are grateful to C. Callan, I. Klebanov, A. Peet and S.-J. Rey for 
useful discussions.  This work was supported in part by a US Department
of Energy Outstanding Junior Investigator Award under grant 
DE-FG02-91ER40671.

\end{document}